\documentclass[12pt,preprint]{aastex}

\shorttitle{The Global Baroclinic Instability in Accretion Disks. II: Linear Analysis.}
\shortauthors{Klahr}

\begin{document}


\title{The Global Baroclinic Instability in Accretion Disks.\\
 II: Local Linear Analysis.}


\author{Hubert Klahr\altaffilmark{1}}
\email{klahr@mpia.de}
\email{ApJ in press}

\affil{Max-Planck-Institut f\"ur Astronomie, Heidelberg, Germany}
\altaffiltext{1}{also at: UCO/Lick Observatory, University of California,
    Santa Cruz, CA 95060}



\begin{abstract}
This paper contains a local linear stability analysis for
accretion disks under the influence of a global radial entropy 
gradient
$\beta = - d \log T / d \log r$ for constant surface density. 
Numerical simulations suggested the existence of
an instability in two- and three-dimensional models of the solar nebula. 
The present paper tries to clarify, quantify, and explain
such a global baroclinic instability for two-dimensional flat 
accretion disk models. As a result linear theory predicts a
transient linear instability that will amplify perturbations
only for a limited time or up to a certain finite amplification.
This can be understood as a result of the growth time of the instability
being longer than the shear time which destroys the modes which are able to grow.
So only non-linear effects can lead to a relevant amplification.
Nevertheless, a lower limit 
on the entropy gradient $\propto\beta \approx  0.22$ for the transient linear instability 
is derived, which can be tested in future non-linear simulations. This would
help to explain the observed instability in numerical simulations as an ultimate
result of the transient linear instability, i.e.\ the Global Baroclinic Instability.
\end{abstract}


\keywords{accretion, accretion disks --- circumstellar matter --- hydrodynamics ---
 instabilities --- turbulence --- methods: numerical --- solar system: formation ---
 planetary systems}
\noindent


\section{Introduction}
In Klahr \& Bodenheimer 2003 (also Klahr \& Bodenheimer 2000) we
presented two-dimensional (and three-dimensional) numerical 
hydrodynamical simulations of accretion disks, without
magnetic fields and self-gravity.
In the case with a radial
entropy $(S)$ gradient,
the disk developed an instability, while in the case of
no radial entropy gradient it stayed perfectly quiet.
\begin{equation}
\frac{d S}{d r} = c_v  \frac{d}{dr} \log P \Sigma^{-\gamma} < 0
\label{entropy_gradient}
\end{equation}
with $c_v$:specific heat, $P$: vertically integrated pressure, $\Sigma$: vertically integrated density and $\gamma$: the adiabatic index ($P \sim \Sigma^\gamma$). 
It is necessary to define a $\gamma$ for the surface density,
because the entire paper will use the two-dimensional
flat approximation of hydrodynamics,
even
it is just a crude approximation of the much more complicated
relation between temperature and surface density. 
Goldreich, Goodman \& Narayan (1986) 
derive $\gamma = (3\Gamma - 1)/(\Gamma + 1)$ for this case,
which relates to $\gamma = 1.354$ for the three-dimensional adiabatic index
$\Gamma = 1.43$ which is realistic for the expected He, H$_2$ mixture
of the solar nebula. 

The non-axisymmetric instability in Klahr \& Bodenheimer (2003) formed waves and grew exponentially at a rate of $\approx \left( 0.1 \Omega\right)$ ( $\Omega =$ orbital
frequency), eventually breaking up into stable vortices.
The formation of vortices is of special interest, because vortices can
be the trigger for a fast and very efficient planet formation process (Klahr 2003).

Measurements of the Reynolds stresses indicated angular momentum transport radially outward 
at an efficiency comparable to $\alpha = 10^{-4} - 10^{-2}$ ($\alpha$ follows from the Shakura-Sunyaev definition $\alpha = \frac{\nu}{c_s H}$, with $\nu$ effective viscosity, $c_s$ speed of sound and $H$ the pressure scale height).
In order to understand this instability a linear
stability analysis is performed in this paper.
The result is a critical value for the entropy gradient, and
the growth rates as they depend on the wave number and time. 

\subsection{Stability and Instability of accretion disks}
Barotropic ($P = P(\Sigma)$) Keplerian disks are centrifugally stable as
predicted by the Rayleigh criterion (see e.g Balbus, Hawley, \&  Stone 1996 
for numerical and analytical investigations).
The situation changes if self gravity (Toomre 1964) or magnetic fields 
become important (Balbus \& Hawley 1991). But both mechanisms are restricted
to parameter ranges unrealistic for a large portion of protoplanetary disks (Gammie 1996).
Therefore, one still looks for alternative effects that could drive turbulence and
thus angular momentum transport in accretion disks. 
R\"udiger, Arlt \& Shalybkov (2002) showed that in vertically stratified 
disks there exists no exponentially growing
linear mode without vertical shear, which was suggested earlier.
The baroclinic instability that is analyzed here in this paper does not
consider the vertical structure explicitly, but only considers a radial entropy gradient.
This might confuse readers who are familiar with the Baroclinic Instability 
as dealt with in geophysical contexts (e.g.\ Pedlosky 1987). There, one
considers the vertical structure to be important. But this is not entirely true,
because one uses a variable vertical variation in density and velocity
for the purpose of introducing a radial pressure gradient in a fluid that
is conveniently assumed to be isothermal and incompressible otherwise. A linear analysis
like the 2-layer model (see Pedlosky 1987) does not consider vertical modes.
Thus, from now on we want to distinguish between the Classical Baroclinic
Instability in the earth atmosphere, which formally arises if the potential 
vorticity gradient changes sign in the vertically stratified atmosphere,
and the Global Baroclinic Instability considered here.
What both instabilities have in common is the radial 
gradient of entropy, which makes the equilibrium state baroclinic (i.e.\ there is
an inclination between pressure and density gradient).
For both instabilities the radial entropy gradient is the {\em sine qua non} condition for instability.
The main difference between the two instabilities is that Keplerian shear is not present in
the Classical Baroclinic Instability. Also, the 2-layer model for
the Classical Baroclinic Instability allows 
one to determine an upper cut-off for the azimuthal wavenumbers which are
unstable; this is possible because the vertical shear kills the instability by eventually
bringing the upper and the lower layers out of phase.
This stabilizing effect of the vertical shear in accretion disks
is not considered in this paper and will be investigated later.
Tentatively, one can speculate that the vertical shear will damp 
high azimuthal wave numbers in disks. Those high wave number modes are also the
strongest damped by the radial shear, as one will see in this paper.
However, the paper focuses on the small to medium range of wave numbers anyway.

   The baroclinic instability in this paper can be regarded as an 
essentially convective instability
in the radial direction but modified by rotation and shear. 
If there would be no shear but only rotation then axisymmetric
perturbations obey the Solberg-H{\o}iland stability criterion.  But
non-axisymmetrically, since the angular momenta of individual fluid
elements is not conserved, the criterion reduces to the Schwarzschild-Ledoux criterion 
$\frac{dP}{dr}\frac{dS}{dr} > 0$ as shown by Cowling(1951). This criterion says that
an instability occurs when radial pressure $P$ and entropy $S$ gradient point in the same
direction. In some sense this paper is an extension of Cowlings work to apply it
for accretion disks by incorporating the treatment of shear.
 
\subsection{Outline of the paper}
Section 2 gives the basic hydrodynamic equations relevant for a
two-dimensional flat accretion disk with a radial entropy gradient. 
In Section 3, a local linear analysis in Lagrangian shear coordinates
is performed. The result is a 4th order dispersion relation, which is a function  of the time-dependent radial wave number $k_x(t)$  and the 
azimuthal wave number $k_y$ in local $x-y$ coordinates. 
Section 4 expands the linear analysis to a global model, but the analysis
is restricted to the radial modes $k_r(t=0) = 0$, showing that the existence
of the non-axisymmetric instability depends on a minimum
entropy gradient. A complete global analysis, as well as the determination of the
eigenfunctions, is left
to future work. Section 5 contains a numerical solution of the linearized equations.
Using the linearized equations from Section 3, it is possible
to give an easy explanation of the special mechanism behind the Global Baroclinic 
Instability. This explanation is given in Section 6. Section 7 summarizes and discusses the results.

\section{Hydrodynamic Equations}
Consider a two-dimensional flat disk in cylindrical coordinates.
The vertically ($z$) integrated density is:
\begin{equation}
\Sigma(r,\phi) = \int_{-\infty}^{+\infty} \rho(r,\phi,z) dz.
\label{Cont_Eq}
\end{equation}
The vertical velocity component ($v_z$) vanishes at infinity, which means that there
is no mass lost or gained from outside.
The governing ideal hydrodynamic equations in polar coordinates are:
\begin{eqnarray}
\partial_t \Sigma + \frac{1}{r}\partial_r \left(r \Sigma v_r\right) + \frac{1}{r}\partial_{\phi} \left(\Sigma v_{\phi}\right) &=& 0,
\label{cont_eq}\\
\partial_t v_r + v_r \partial_r v_r + \frac{v_\phi}{r}\partial_{\phi} v_{r} &=& - \frac{1}{\Sigma}\partial_r P + \frac{v_{\phi}^2}{r} - \Omega_{kep}^2 r,
\label{rad_mom_eq}\\
\partial_t v_\phi + v_r \partial_r v_\phi + \frac{v_\phi}{r}\partial_{\phi} v_{\phi} + v_r \frac{v_\phi}{r} &=& - \frac{1}{r\Sigma}\partial_\phi P,
\label{azi_mom_eq}\\
\partial_t P + v_r\partial_r P + \frac{v_\phi}{r}\partial_{\phi} P  &=& -\gamma P \left(\frac{1}{r}\partial_r r v_r + \frac{1}{r}\partial_\phi v_\phi \right)\label{ene_eq},
\end{eqnarray}
where $P$ is the vertically integrated pressure, 
\begin{equation}
P(r,\phi) = \int_{-\infty}^{+\infty} p(r,\phi,z) dz,
\end{equation}
and $\Omega_{kep}$ is the local Keplerian frequency.
The energy equation,
\begin{equation}
\partial_t E_{th} + \frac{1}{r}\partial_r \left(r v_r E_{th}\right) + \frac{1}{r}\partial_{\phi} \left(v_{\phi} E_{th}\right) = -P \nabla v,
\end{equation}
is replaced by an equation for the pressure, $E_{th} = P / (\gamma -1)$, via the ideal gas equation with adiabatic index $\gamma$. 
The baroclinicity of a general flow is given by 
the baroclinic term: 
\begin{equation}
\nabla \rho(r,z,\phi) \times \nabla p(r,z,\phi) \neq 0.
\end{equation}
In a three-dimensional but axisymmetric equilibrium state of a disk, 
there is in fact a non-zero baroclinic term above and below the midplane, which vanishes at the midplane
itself ($z = 0$):
\begin{equation}
\nabla \rho(r,z) \times \nabla p(r,z) \neq 0.
\end{equation}
In the two-dimensional flat approximation used in this paper, this term vanishes
for the axisymmetric mean state,
\begin{equation}
\nabla \Sigma(r) \times \nabla P(r) = 0,
\end{equation}
but still
the disk is baroclinic. This is shown by the existence of the two independent equations 
for density and internal energy, reflecting the inclination of
the equipressure surface towards the equidensity surface with radius.
Again, most important for the instability is that the non-axisymmetric deviations from the 
mean state can lead to the rise of the baroclinic term even in two dimensions,
\begin{equation}
\nabla \Sigma(r,\phi) \times \nabla P(r,\phi) \neq 0,
\end{equation}
and vorticity can be generated.
\subsection{Defining the mean state}
The quantities $Q = (\Sigma, v_r, v_\phi, P)$ can be split into
a time independent axisymmetric mean part $\bar Q$ and a perturbation $Q^\prime$ as
\begin{equation}
Q = \bar Q(r) + Q(r,\phi,t)^\prime
\label{pert_eq}
\end{equation}
The constant density background is assumed to follow a simple power law:
\begin{equation}
\bar\Sigma(r) = \Sigma_0 \left(\frac{r}{r_0}\right)^{-\beta_\Sigma}.
\label{sigma_dist}
\end{equation}
The pressure is distributed in the same fashion:
\begin{equation}
\bar P(r) = P_0 \left(\frac{r}{r_0}\right)^{-\beta},
\label{press_dist}
\end{equation}
thus for $\beta_S$ the entropy gradient (Eq.\ \ref{entropy_gradient}) it follows:
\begin{equation}
\partial_r S = c_v \frac{-\beta +\gamma \beta_\Sigma}{r}\,\, {\rm and}\,\, \beta_S = \beta - \gamma \beta_\Sigma  
\label{entro_dist}
\end{equation}
There is no radial mean flow, which means:
\begin{equation}
\bar{v_r} = 0.
\end{equation}
The mean rotation is axisymmetric and determined by the radial component of the Euler equation, thus
the equilibrium between radial pressure gradient, gravity and
centrifugal acceleration is:
\begin{equation}
0 = - \frac{1}{\bar\Sigma}\partial_r \bar P + \frac{{\bar{v_{\phi}}^2}}{r} - \Omega_{kep}^2 r,
\end{equation}
which leads to the definition for the local orbital frequency:
\begin{equation}
\Omega = \frac{\bar{v_{\phi}}}{r} = \sqrt{\Omega_{kep}^2 + \frac{1}{r \bar\Sigma}\partial_r \bar P }.
\end{equation}
Using the definitions from Eqs.{\ref{sigma_dist}} and {\ref{press_dist}} one gets:
\begin{equation}
\Omega  = \sqrt{\Omega_{kep}^2 - \beta \frac{P_0}{\Sigma_0 r_0^2}\left(\frac{r}{r_0}\right)^{\beta_\Sigma - \beta - 2}}.
\end{equation}
In the case of no radial pressure gradient
\begin{equation}
\Omega =  \Omega_{kep}.
\end{equation}
Unfortunately, $\Omega$ does not in general follow a power law; but,
as $\Omega_{kep}^2 \sim r^{-3}$, it is possible for $\beta_\Sigma - \beta -2 = -3$ to define an $\Omega_0$
\begin{equation}
\Omega =  \Omega_0 \left(\frac{r}{r_0}\right)^{-\frac{3}{2}},
\end{equation}
with
\begin{equation}
\Omega_0 = \sqrt{\Omega_{kep}^2 - \frac{P_0}{\Sigma_0 r_0^2}} = \Omega_{kep}\sqrt{1 - \frac{H^2}{\gamma r_0^2}}.
\end{equation}
This situation is for instance given if the surface density is constant over radius ($\beta_\Sigma = 0$)
and the sound speed $c_s$ is a constant fraction of the Keplerian velocity; e.g., the relative pressure scale
height $H/r$ is constant with radius. This leads to the definition of two cases, in which
A: there is no radial pressure gradient $(\beta = 0)$ or B: a pressure gradient exists
with $\beta = 1$. For all cases of our analytic treatment, we simplify the 
density structure to be $\beta_\Sigma = 0$ $(\bar\Sigma = \Sigma_0)$.
In case A the rotational frequency is exactly Keplerian, but in case B 
the disk is slightly sub-Keplerian while still maintaining a Keplerian $(\sim r^{-1.5})$ profile.

%
%

\section{Local Linear Analysis}
This linear analysis will demonstrate the dependence of the 
instability on the existence of non-axisymmetric modes
and the radial entropy gradient. In contrast to the
WKB-approximation for tightly wound spirals in self gravitating disks,
the baroclinic instability disappears for radial wave numbers much larger 
than the azimuthal wave numbers ($k_x(t) \gg k_y$). This provides a
potential problem for the instability in a shear flow.
The radial wave numbers always tend to increase because of the winding up
of spirals.
In a WKB-approximation, one would immediately erase all baroclinic unstable
modes. Thus, we can not make use of this simplification.

Non-axisymmetric perturbations cannot have a simple waveform (Goldreich \& Lynden-Bell 1965) because
of the effect of the shearing background on the wave crests. Thus one adopts Lagrangian shearing coordinates
\begin{eqnarray}
 r^\prime &=& r ,\\
\phi^\prime &=& \phi  - \Omega(r) t,\\
t^\prime &=& t,
\end{eqnarray}
comoving with the unperturbed shear flow. Then, it follows that,
\begin{eqnarray}
 \partial_r &=& \partial_{r^\prime} - t \frac{d\Omega}{dr}\partial_{\phi\prime},\\
\partial_\phi &=& \partial_{\phi^\prime}.
\end{eqnarray}
The new time derivative is defined as:
\begin{equation}
\partial_t=\partial_{t^\prime} - \Omega(r) \partial_\phi.
\end{equation}
The azimuthal velocity is also split into two parts:
\begin{eqnarray}
v_\phi &=& v_{\phi^\prime} + \Omega(r) r.
\end{eqnarray}
Linear perturbations are assumed to have the space dependence $e^{i(k^\prime_r r^\prime + m \phi^\prime)}$
in Local Lagrangian coordinates.
Thus, one has now a time dependence in the radial wave number with respect to the
non shearing one:
\begin{equation}
k_r \leftarrow k_r (t) = k^\prime_r - m t \frac{d\Omega}{dr} = k^\prime_r + q \frac{m}{r} \Omega t,
\end{equation}
using $\Omega \propto r^{-q}$. 
One starts from the vertically integrated hydrodynamic 
equations in cylindrical coordinates, performs
a linearisation, and solves the linear equations for
the frequencies $\omega$, and the growth-rate $\Gamma = - i \omega$.

In Lagrangian shear coordinates (dropping the primes at all coordinates), the Eqs.\ (\ref{cont_eq}) to (\ref{ene_eq}) transform to:
\begin{eqnarray}
\partial_t \Sigma &+& \frac{1}{r}\left(\partial_r - t \frac{d\Omega}{dr}\partial_\phi\right) \left(r \Sigma v_r\right)  + \frac{1}{r}\partial_{\phi} \Sigma v_{\phi} = 0,
\label{lcont_eq}\\
\partial_t v_r &+& v_r \left(\partial_r - t \frac{d\Omega}{dr}\partial_\phi\right) v_r+ \frac{v_\phi}{r}\partial_{\phi} v_{r} = - \frac{1}{\Sigma}\left(\partial_r - t \frac{d\Omega}{dr}\partial_\phi\right) P + \frac{\left(\Omega r + v_{\phi}\right)^2}{r} - \Omega_{kep}^2 r,
\label{lrad_mom_eq}\\
\partial_t v_\phi &+& v_r \left(\partial_r - t \frac{d\Omega}{dr}\partial_\phi\right) \left(v_\phi + \Omega r\right) + \frac{v_\phi}{r}\partial_{\phi} v_{\phi}+ v_r \frac{\left(v_\phi + \Omega r\right)}{r}   = - \frac{1}{r\Sigma}\partial_\phi P,
\label{lazi_mom_eq}\\
\partial_t P &+& v_r\left(\partial_r- t \frac{d\Omega}{dr}\partial_\phi\right) P+ \frac{v_\phi}{r}\partial_{\phi} P  = -\gamma P \left(\frac{1}{r}\left(\partial_r- t \frac{d\Omega}{dr}\partial_\phi\right) r v_r + \frac{1}{r}\partial_\phi v_\phi \right)\label{lene_eq}.
\end{eqnarray}
Using a local approximation
simplifies the basic equations.
Assuming $r = R + x$, $\phi = y / R$, $v_\phi = \Omega_0 R + v_y$ and $x \ll R$, the ideal hydrodynamical equations are in a frame corotating with the mean orbital frequency $\Omega_0$:
\begin{eqnarray}
\partial_t \Sigma &+& (\partial_x + q \Omega_0 t \partial_y) \Sigma v_x +   \partial_y\Sigma v_y = 0,\\
\partial_t v_x &+& v_x (\partial_x + q\Omega_0 t \partial_y) v_x  + v_y \partial_y v_x= - \frac{1}{\Sigma}(\partial_x + q \Omega_0 t  \partial_y) P + 2 \Omega_0 v_y - 2 q \Omega_0 x,\\
\partial_t v_y &+& v_x (\partial_x + q \Omega_0 t \partial_y) v_y + v_x \Omega_0 + v_x\partial_r \Omega R + v_y \partial_y v_y= - \frac{1}{\Sigma}\partial_y P,\\
\partial_t P &+& v_x \left(\partial_x  + q \Omega_0 t \partial_y\right) P  + v_y \partial_y P = -\gamma P \left( \left(\partial_x + q \Omega_0 t \partial_y \right) v_x + \partial_y v_y \right).
\end{eqnarray}
These equations are linearized, and using Eq.\ref{pert_eq} and the definition of Oort's constant (see Binney \& Tremaine 1987),
\begin{equation}
B(r) = -\frac{1}{2}\left[\partial_r\left(\Omega r \right)+ \Omega \right] = \left(\frac{q}{2}-1\right) \Omega,
\end{equation}
it follows (dropping the index $_0$ at $\Omega_0$ in the following) that
\begin{eqnarray}
\partial_t \Sigma^\prime + (\partial_x + q \Omega t \partial_y) \Sigma_0 v_x^\prime +  \Sigma_0 \partial_y v_y^\prime &=& 0,\\
\partial_t v_x^\prime &=& - \frac{1}{\Sigma_0}(\partial_x + q \Omega t  \partial_y) P^\prime + \frac{\Sigma^\prime}{\Sigma_0^2}\partial_x \bar{P} + 2 \Omega v_y^\prime,\\
\partial_t v_y^\prime  - 2 B v_x^\prime &=& - \frac{1}{\Sigma_0}\partial_y P^\prime,\\
\partial_t P^\prime + v_x^\prime \partial_x \bar{P} &=& -\gamma \bar{P} \left( \left(\partial_x + q \Omega t \partial_y \right) v_x^\prime + \partial_y v_y^\prime \right).
\end{eqnarray}
The radial pressure gradient is given by 
$\partial_x \bar P = -\frac{\beta}{R} \bar P$.
This simple model helps to
understand the basic mechanism, and its flaws will
be overcome by the global model.
Indeed, the global analysis delivers a limit
on the pressure gradient $\beta$ while
this local model only restricts $\beta \ne 0$.

Using the quantity $B$ gives the possibility to keep the
analysis general for arbitrary rotational profiles (e.g.\ Keplerian rotation, rigid rotation or Rayleigh unstable flows).

For the perturbed state we assume:
\begin{eqnarray}
Q(x,y,t) = Q_a e^{-i\left( \omega t - k_y y - k_x(t) x \right)},
\end{eqnarray}
even complex
exponentials of the form assumed for $Q(x,y,t)$ are not exact solutions to 
(41)-(44), and a priori
one would not expect them to be good approximations unless the
dimensionless shear rate $q$ is small. 
The point is, that $\omega$ generally is also a function of time $\omega(t)$.
On the other hand, following
Goldreich and Lynden-Bell (1965), Balbus and Hawley (1992), Ryu and
Goodman (1992), and many others [ultimately this technique goes back
to Kelvin at least], one can derive from Eqs.\ (41)-(44) a set of
four ordinary differential equations in time for the functions
$Q_a(t)$ defined by:
\begin{eqnarray}
Q(x,y,t)=Q_a(t) e^{i(k_x(0)+iq\Omega t k_y)x +ik_y y}.
\end{eqnarray}
The behavior of $Q_a(t)$ is not sinusoidal except approximately so 
at  large
$|t|$, where it can be decomposed into a sum of terms with slowly  
changing WKB frequencies and amplitudes.
In this way one can calculate the correct swing amplification  
factor and
compare with the approximate results in the second panel of fig.\ 5.
This numerical check was carried out and is described in section 5. 
The results show that the approximate solution Eq.\ (45) leads to
a qualitatively correct result in the analytic approach.

Using Eq.\ (45) to search for eigen-values of Eqs.\ (41)-(44) one
obtains:
\begin{eqnarray}
-i \omega \frac{\Sigma_a}{\Sigma_0} + i \left(k_x + q k_y \Omega t\right)  v_{xa} + i k_y v_{ya} &=& 0, \label{eq_lina1}\\
\frac{\beta \bar P}{R \Sigma_0} \frac{\Sigma_a}{\Sigma_0} - i \omega  v_{xa}  - 2 \Omega v_{y a} + i \left(k_x + q k_y\Omega t\right) \frac{P_a}{\Sigma_0}  &=& 0,\\
- 2 B v_{xa} - i \omega  v_{y a}  + i k_y\frac{P_a}{\Sigma_0} &=& 0,\\
\left(i \left(k_x + q k_y\Omega t\right) \gamma - \frac{\beta}{R}\right)  \bar \frac{P}{\Sigma_0} v_{xa} + i k_y \gamma \bar \frac{P}{\Sigma_0}  v_{y a} - i\omega \frac{P_a}{\Sigma_0}   &=& \label{eq_lina2} 0
\end{eqnarray}
The two terms proportional to $\beta$ make all the difference with
respect to the stability analysis found in the literature (e.g.\ Binney \& Tremaine 1987).
Without them one retains only the known stability criteria, because
setting $\beta = 0$ produces a third order system out of this fourth order
system. With $\beta = 0$ surface density $\Sigma$ and pressure $P$ are no
longer independent.

The mean pressure can be replaced by the expression for the sound speed as:
\begin{equation}
\gamma \frac{\bar {P}(x)}{\Sigma_0} = c_s^2(x).
\end{equation}
The linear modal Eqs.\ (\ref{eq_lina1} - \ref{eq_lina2}) permit solutions for
$\Gamma = - i \omega$ only for a vanishing determinant:
\begin{mathletters}
\begin{equation}
\left|
\begin{array}{cccc}
\Gamma & i \left(k_x + q k_y\Omega t\right) & i k_y & 0 \\
\frac{\beta c_s^2}{R \gamma} & \Gamma &  - 2 \Omega & i \left(k_x + q k_y\Omega t\right) \\
0 & - 2 B & \Gamma & i k_y \\
0 & c_s^2  \left( i \left(k_x + q k_y\Omega t\right) - \frac{\beta} {R \gamma}\right) & i c_s^2 k_y & \Gamma
\end{array} \; \; \right|\;
\;  \;
\; = 0 \;
\end{equation}
\end{mathletters}
One obtains the dispersion relation:
\begin{eqnarray}
\Gamma^4 &+& \Gamma^2 \left[\left(k_x^2 + 2 k_x k_y q \Omega t+ k_y^2 (1 + q^2 \Omega^2 t^2)\right) c_s^2 + \kappa^2\right] \nonumber \\
&+& 2 \Gamma \left[\frac{q}{2}\Omega \left(k_x k_y + k_y^2 q \Omega t\right)c_s^2 + i k_y \frac{\beta c_s^2}{R \gamma}\left(2-\frac{q}{2}\right) \Omega\right]\nonumber\\
&-& \frac{\beta^2 c_s^4 k_y^2}{R^2 \gamma^2} = 0.
\end{eqnarray}

If one uses the dimensionless azimuthal wave number $m = k_y R$ and the
time-dependent dimensionless radial wave number $n(t) = k_x R + q m \Omega t$ this relation 
simplifies to:
\begin{eqnarray}
\Gamma^4 &+& \Gamma^2 \left[\left(n^2 + m^2\right)\left(\frac{H}{R}\right)^2  + \frac{\kappa^2}{\Omega^2}\right]\Omega^2 \nonumber \\
&+& 2 \Gamma \left[\frac{q}{2} \left(m n\right) + i m\frac{\beta}{\gamma}\left(2-\frac{q}{2}\right)\right]\left(\frac{H}{R}\right)^2 \Omega^3\nonumber\\
&-& \frac{\beta^2 m^2}{\gamma^2} \left(\frac{H}{R}\right)^4 \Omega^4 = 0.
\end{eqnarray}
This relation is not restricted to Keplerian disks but valid for all centrifugally
supported rotating bodies with the general description for $q$.
In the case of a Keplerian disk with $q = 1.5$ it is
\begin{eqnarray}
\Gamma^4 &+& \Gamma^2 \left[\left(n^2 + m^2\right)\left(\frac{H}{R}\right)^2  + 1\right]\Omega^2 \nonumber \\
&+& 2 \Gamma \left[\frac{3}{4} \left(m n\right) + i m\frac{\beta}{\gamma} \frac{5}{4} \right]\left(\frac{H}{R}\right)^2 \Omega^3\nonumber\\
&-& \frac{\beta^2 m^2}{\gamma^2} \left(\frac{H}{R}\right)^4 \Omega^4 = 0.
\end{eqnarray}

This 4th order equation has two high and two low frequency solutions.
In Fig.\ 1 the growth-rate $Re(\Gamma)$ and frequencies $Im(\Gamma)$ are
plotted for the case $H/R = 0.1$, $m = 10$, $\beta = 1$ in units of the rotational
frequency $\Omega$ as a function of the time-dependent radial wave number $n(t)$.
The roots were found using the IDL routine FZ\_ROOTS.
One recognizes in fact two solutions with frequencies close to zero
and two solutions with frequencies larger than the rotational frequency 
$\Omega$. The high frequency solutions are the ones which show anti-symmetric
behavior in growth-rates with respect to time. Those are the ones which are negative
for $n<0$.

The dispersion relation can be more easily interpreted by considering the high
frequency and low frequency limits separately.
\subsection{High Frequency Waves}
For the high frequency part $|\Gamma| \gg \Omega$ one may neglect the constant term but not the term linear in $\Gamma$. This is a depressed cubic equation and its solution is given by Cardano's method,
which shall not be done here.
For the $m=0$ case one has a
2nd order equation
\begin{eqnarray}
\Gamma_h^2  &=& - \left[n^2\left(\frac{H}{R}\right)^2  + \frac{\kappa^2}{\Omega^2}\right]\Omega^2 .
\end{eqnarray}
This axisymmetric solution is the usual dispersion relation of sound waves, with the epicyclic frequency $\kappa$,
and there is no growth unless $\kappa^2 < 0 => q > 2$, which is known as Rayleigh's stability criterion.
\subsection{Low Frequency Waves}
Considering the low frequency $|\Gamma| \ll \Omega$ part of the dispersion relation
one neglects the highest order $\Gamma^4$, thus retaining a second order system
for the low frequency waves only:
\begin{eqnarray}
& & \Gamma_l^2 \left[\left(n^2 + m^2\right)  + \frac{\kappa^2}{\Omega^2}\left(\frac{H}{R}\right)^{-2}\right] \nonumber \\
&+& 2 \Gamma \left[\frac{q}{2} \left(m n\right) + i m\frac{\beta}{\gamma}\left(2-\frac{q}{2}\right)\right] \Omega\nonumber\\
&-& \frac{\beta^2 m^2}{\gamma^2} \left(\frac{H}{R}\right)^2 \Omega^2 = 0,
\end{eqnarray}
which is a quadratic equation with the roots:
\begin{eqnarray}
\Gamma_l &=& - m \frac{\frac{q}{2} n + i \frac{\beta}{\gamma}\left(2-\frac{q}{2}\right)}{n^2 + m^2 + \frac{\kappa^2}{\Omega^2}\frac{R^2}{H^2}} \Omega \nonumber\\
&\pm& \frac{\beta}{\gamma} m \frac{H}{R} \Omega \frac{\sqrt{\left[ \left(\frac{R \gamma q}{2 H \beta}\right)^2+1\right] n^2 + m^2 + i \left(2-\frac{q}{2}\right)\frac{R^2 \gamma q}{H^2 \beta} n   - \frac{R^2}{H^2}\frac{q^2}{4}}}{n^2 + m^2 + \frac{\kappa^2}{\Omega^2}\frac{R^2}{H^2}}.
\end{eqnarray}

In the barotropic case $\beta = 0$ there is only one solution, which is:
\begin{eqnarray}
\Gamma_{l0} &=& - m \frac{q n}{n^2 + m^2 + \frac{\kappa^2}{\Omega^2}\frac{R^2}{H^2}} \Omega.
\end{eqnarray}
This is not to be confused with an instability, because it is antisymmetric
with respect to $n(t)$. This means there is no effective growth over a period of time:
\begin{eqnarray}
\int_{-t_0}^{t_0}\Gamma_{l0}(t) dt &=& 0.
\end{eqnarray}

For a baroclinic $\beta \ne 0$ rigid rotating object ($q = 0$) (e.g.\ a star) the 
growth will stay linear for all times, because 
$n$ is no longer time-dependent, i.e.\ there is no shear which winds up the spirals:
\begin{eqnarray}
\Gamma_l &=& - m \frac{2 i \frac{\beta}{\gamma}}{n^2 + m^2 + 4 \frac{R^2}{H^2}} \Omega \nonumber\\
&\pm& \frac{\beta}{\gamma} m \frac{H}{R} \Omega \frac{\sqrt{n^2 + m^2 }}{n^2 + m^2 + 4\frac{R^2}{H^2}}.
\end{eqnarray}
This equation should be appropriate to describe baroclinic modes
in centrifugally supported stars and other fast rotating objects.

For a Keplerian disk ($q = 3/2$) one has nevertheless time-dependent radial
wave numbers $n(t)$:
\begin{eqnarray}
\Gamma_l &=& - m \frac{\frac{3}{4} n(t) + i \frac{5}{4}\frac{\beta}{\gamma}}{n(t)^2 + m^2 + \frac{\kappa^2}{\Omega^2}\frac{R^2}{H^2}} \Omega \nonumber\\
&\pm& \frac{\beta}{\gamma} m \frac{H}{R} \Omega \frac{\sqrt{\left[ \left(\frac{3 R \gamma}{4 H \beta}\right)^2+1\right] n(t)^2 + m^2 + i \frac{15}{8}\frac{R^2 \gamma}{H^2 \beta} n(t)   - \frac{33}{16} \frac{R^2}{H^2}}}{n(t)^2 + m^2 + \frac{\kappa^2}{\Omega^2}\frac{R^2}{H^2}}.
\end{eqnarray}
In the barotropic case $\beta = 0$ most terms vanish and the growth-rates are 
\begin{eqnarray}
\Gamma_0(t) &=& - \frac{\frac{3}{2} m n(t)}{\left(n(t)^2 + m^2\right)  + \frac{R^2}{H^2}} \Omega.
\end{eqnarray}
What is described here, is how a pattern with leading spirals gets unwound
and its amplitude increases over this period. As soon as the spiral
is open, and the winding up to trailing waves starts, the amplitude
decreases again for the same amount it initially was increased.
This solution has no imaginary part which means it is completely in
phase with the local rotation. This dependence of $\Gamma_0$
on time (respectively time dependent wave number) is shown in Fig.\ 2. and
Fig.\ 3 as the dotted lines. Fig.\ 2 and 3 only differ for the 
range of radial wave numbers plotted. The chosen parameters are
$H/R = 0.055$, and $m = 48$ 
The anti-symmetry between negative
and positive wave numbers $n$ tells us that here effectively no 
amplification is taking place.

In the baroclinic case $\beta \ne 0$  a bifurcation of
the solution branch occurs. One gets two different waves with growth-rates $Re(\Gamma^+)$ and $Re(\Gamma^-)$ and with two different frequencies ($Im(\Gamma^+),Im(\Gamma^-)$, which resemble the phase velocity of the different eigenfunctions.
The $\Gamma^+$ solution is plotted as a solid line in Fig.\ 2 and 3 for $\beta = 1$.
For large negative values of $n$, which corresponds to tightly wound
leading waves, a behavior close to the barotropic solution occurs.
But as the spiral wave approaches a state where it is basically open
the amplification does not go to zero but remains small but positive for all times.
During the up winding of trailing waves there is no decay but even a
further transient amplification of the spiral pattern.

The other solution branch $\Gamma^-$, which is given by the dashed line,
has zero growth as it starts with negative wave numbers and 
even decays when it approaches the unwound spiral $n(t)=0$ state (i.e. the
respective time). As the spiral winds up to the trailing state this pattern
has a decaying amplitude and is thus of no relevance for
the global baroclinic instability.

The two solutions have different eigenfunctions with different
phase velocities and growth behavior. The important effect
in the global baroclinic instability is the bifurcation of the 
solution branch into a transiently linear growing mode 
and a monotonically linear decaying mode. The idea is now that
the transient linear growth can lead to the development of
non-linear turbulence including vortices and explain the observed instability
in the non-linear simulations of Klahr \& Bodenheimer (2003).

Hence, this shows that our vertically integrated baroclinic models have 
no pure linear unstable modes in the presence of shear. 
The instability appears to be transient and one has to investigate mode
coupling or finite amplitude effects in a non linear
analysis. In the case of rigid rotation one finds nevertheless
a true linear instability. This might be of importance for rigid rotating
objects like stars close to their breakup velocity.

The local analysis gives no minimum value for the entropy
gradient or, correspondingly, $\beta$.
Any $\beta \ne 0$ (positive and negative!) leads to transient growth $\sim |\beta|$.
But with a global analysis the situation changes.

\section{Restricted Global Linear Analysis}
It is not possible to perform a general global analysis with simple $k$ Fourier modes
in the radial direction ($k$ independent on $r$). 
Thus the analysis will be restricted to a special family of
solutions which are described by
\begin{eqnarray}
Q(r,\phi,t)&=& 
Re[Q_a e^{-i\left(\omega t - m \phi - q m \Omega(r) t\right)}].
\end{eqnarray}
The assumption $\partial_t \omega << 1$ is the same as we used 
in previous section. 
These are spirals which are open at the time $t = 0$, leading for $t<0$ and 
trailing for $t>0$.
%
Splitting the quantities in the global hydrodynamic equations (Eq.\ \ref{lcont_eq}-\ref{lene_eq}) 
in Lagrangian coordinates into a mean and a fluctuating part, without doing the local approximation 
and dropping terms of higher order, leads to the global linearized equations:
\begin{eqnarray}
\partial_t \Sigma^\prime + \frac{1}{r}\left(\partial_r + \frac{q}{r} \Omega t \partial_\phi\right) \left(r \Sigma_0 v_r^\prime\right)  + \frac{\Sigma_0}{r}\partial_{\phi}  v_{\phi}^\prime &=& 0,
\label{glcont_eq}\\
\partial_t v_r^\prime &=& - \frac{1}{\Sigma_0}\left(\partial_r + \frac{q}{r} \Omega t\partial_\phi\right) P^\prime - \frac{\beta}{r \Sigma_0^2}  \bar P \Sigma^\prime + 2 v_{\phi}^\prime \Omega, \label{glrad_mom_eq}\\
\partial_t v_\phi^\prime +  - 2 B v_r^\prime  &=& - \frac{1}{r \Sigma_0}\partial_\phi P^\prime,
\label{glazi_mom_eq}\\
\partial_t P^\prime - v_r ^\prime \frac{\beta \bar P}{r}  &=& -\gamma \bar P \left(\frac{1}{r}\left(\partial_r + \frac{q}{r} \Omega t \partial_\phi\right) r v_r^\prime + \frac{1}{r}\partial_\phi v_\phi^\prime  \right)\label{glene_eq},
\end{eqnarray}
Solution of the linearized equations can be written as a sum of terms of the form of $Q^\prime = Q(r^\prime)_a e^{-i\left(\omega t^\prime - m \phi^\prime \right)} $.
Without a radial variation of the initial perturbation  $Q(r,\phi,t)$ at (t=0) (i.e.\ $\partial_r Q_a = 0$)
one has for the radial derivative:
\begin{eqnarray}
-i \omega \frac{\Sigma_a}{\Sigma_0} + \left(\frac{1}{r} + i \frac{m}{r} q \Omega t\right) v_{ra}  + i \frac{m}{r} v_{\phi a} &=& 0,
\label{cont_G}\\
\frac{\beta}{r \Sigma^2} \bar P \Sigma_a -i \omega  v_{ra}  - 2  \Omega v_{\phi a}+ \frac{i n}{r \Sigma_0} P_a&=& 0, \label{rad_mom_G}\\
 + \frac{i m}{\Sigma_0} q \Omega t P_a - 2 B v_{ra} -i \omega v_{\phi a}  + \frac{i m}{r \Sigma_0} P_a &=& 0,
\label{azi_mom_G}\\
\frac{\bar P}{r}\left(\gamma \left(1 + i m q \Omega t\right) - \beta \right) v_{ra}  + \gamma \frac{i m \bar P}{r} v_{\phi a} -i \omega P_a &=& 0\label{energ_G},
\end{eqnarray}
%
These four equations (Eq.(\ref{cont_G}),(\ref{rad_mom_G}),(\ref{azi_mom_G}), and (\ref{energ_G})) constitute a linear system of equations with $\Gamma = - i \omega$ and using the definition:
\begin{equation}
n(r) = m q \Omega(r) t
\end{equation}
this can be written as:
\begin{mathletters}
\begin{equation}
\left(
\begin{array}{cccc}
\Gamma & \frac{1+ i n}{r} & i \frac{m}{r} & 0 \\
\frac{\beta c_s^2}{r \gamma} & \Gamma &  - 2 \Omega & i \frac{n}{r} \\
0 & - 2 B & \Gamma & \frac{i m}{r} \\
0 &  \frac{1 + i n - \frac{\beta} {\gamma}}{r}c_s^2 & \frac{i m}{r}c_s^2 & \Gamma
\end{array} \; \; \right)\;
\; \left(
\begin{array}{c}
 \frac{\Sigma_a}{\Sigma_0}\\
 v_{ra}\\
 v_{\phi a}\\
 \frac{P_a}{\Sigma_0}
\end{array} \; \; \right) \;
\; = \left(
\begin{array}{c}
 0\\
 0\\
 0\\
 0
\end{array} \; \; \right) \;
\end{equation}
\end{mathletters}
For non trivial solutions the determinant of the coefficient matrix must vanish.
Thus the following dispersion relation is obtained, using $B = \left(\frac{q}{2}-1\right) \Omega$:
\begin{eqnarray}
\Gamma^4 &+& \Gamma^2 \left[\frac{m^2 + n^2 - in - \frac{\beta}{\gamma}}{r^2} c_s^2 - 4\left(\frac{q}{2}-1\right) \Omega^2\right] \nonumber \\
&+& \Gamma \left[2 m n \frac{q}{2}\Omega \frac{c_s^2}{r^2} + 2 i m \Omega\left(\left(2 - \frac{q}{2}\right)\frac{\beta}{\gamma} -1 \right)\frac{c_s^2}{r^2}\right]\nonumber\\
&-& \frac{\beta^2 c_s^4 m^2}{r^4 \gamma^2} = 0.
\end{eqnarray}
A numerical method like FZ\_ROOTS from IDL allows us to investigate the
entire parameter space ($q,n,m,\beta,H/R$) for this general dispersion relation.
The results for $m=60$, $q=1.5$, $\beta = 1$ are plotted in 
Fig.\ 4. One recognizes just like in the local case two high
and two low frequency solutions. The high frequency ones have an 
anti-symmetric behavior with respect to time while the low frequency ones
are not at all symmetric in time.
The interested reader is invited to explore this parameter space on his/her own.

For an accretion disk with $q=1.5$ the complete dispersion relation is: 
\begin{eqnarray}
\Gamma^4 &+& \Gamma^2 \left[m^2 + n^2 - in - \frac{\beta}{\gamma} + \left(\frac{H}{R}\right)^{-2}\right]\left(\Omega\frac{H}{R}\right)^2 \nonumber \\
&+& \Gamma  \left[-\frac{3}{2} m n + 2 i m \left(\frac{5\beta}{4\gamma} -1 \right)\right]\Omega^3 \left(\frac{H}{R}\right)^2\nonumber\\
&-& \frac{\beta^2 }{\gamma^2} m^2 \left(\Omega \frac{H}{R}\right)^4  = 0.
\end{eqnarray}
If the last term vanishes because $\beta =0$, one retains a third order equation which leads to the Rayleigh stability criterion.
This expression looks very unhandy but one can again split the solution into a high and a low
frequency solution for $\Gamma$.

\subsection{High Frequency Waves}
For the high frequency part $|\Gamma| \gg \Omega$ one neglects again the constant and the linear term in $\Gamma$
\begin{eqnarray}
\Gamma_h^2 &=& - \frac{m^2 + n^2 - in - \frac{\beta}{\gamma}}{r^2} c_s^2 + 4\left(\frac{q}{2}-1\right) \Omega^2,
\end{eqnarray}
which is for $m=0$ basically nothing else than a Solberg-H{\o}iland criterion for stability
against axisymmetric perturbations:
\begin{equation}
0 > g\partial_r \left(ln P\right) \left(\nabla - \nabla_{ad}\right)  -\frac{1}{r^3}\partial_r \left(r^4 \Omega^2 \right)
\end{equation}

For instability,  $\beta/\gamma$ has to be steeper than $\left(R/H\right)^2$ which is
in a usual thin accretion disk probably never the case (e.g for a reasonable $\frac{H}{R} \simeq 0.1$
one would need a $\beta > 100$ to make the flow unstable).
Such an instability may become important if the rotational profile becomes
steeper because then $\kappa$ drops.

\subsection{Low Frequency Waves}

Considering the low frequency part of the dispersion relation
one neglects $\Gamma_l^4 \ll 1$, thus
\begin{eqnarray}
& & \Gamma_l^2 \left[m^2 + n^2 - in - \frac{\beta}{\gamma} - 4\left(\frac{q}{2}-1\right)\left(\frac{H}{R}\right)^{-2} \right]\nonumber \\
&+& \Gamma_l \left[2 m n \frac{q}{2} + 2 i m \left(\left(2 - \frac{q}{2}\right)\frac{\beta}{\gamma} -1 \right)\right]\Omega\nonumber\\
&-& \frac{\beta^2 m^2}{\gamma^2}\left(\frac{H}{R}\right)^{2}\Omega^2 = 0.\end{eqnarray}
which can be solved for $\Gamma_l$
defining
\begin{eqnarray}
C_2 &=&\left[m^2 + n^2 - in - \frac{\beta}{\gamma} - 4\left(\frac{q}{2}-1\right)\left(\frac{H}{R}\right)^{-2} \right], \\
C_1 &=& \left[m n \frac{q}{2} + i m \left(\left(2 - \frac{q}{2}\right)\frac{\beta}{\gamma} -1 \right)\right],\\
C_0 &=& \frac{\beta^2 m^2}{\gamma^2}\left(\frac{H}{R}\right)^{2}.
\end{eqnarray}
This leads to the quadratic equation
\begin{eqnarray}
\Gamma_l &=& \left[\frac{C_1}{C_2} \pm \frac{\sqrt{C_1^2 + C_0 C_2}}{C_2}\right] \Omega,
\end{eqnarray}
The complete solution for Keplerian disks $q=1.5$ is:
\begin{eqnarray}
\Gamma_l &=& m \frac{\frac{3}{4} n  + i \left(\frac{5}{4}\frac{\beta}{\gamma} - 1 \right)}{ m^2 + n^2 - in - \frac{\beta}{\gamma} + \frac{R^2}{H^2}}\Omega\\
&\pm& m \frac{\sqrt{\left(\frac{3}{4} n \right)^2 + i \frac{3}{2} n \left(\frac{5}{4}\frac{\beta}{\gamma} - 1 \right) - \left(\frac{5}{4}\frac{\beta}{\gamma} - 1 \right)^2 + \frac{\beta^2}{\gamma^2}\left(\frac{H}{R}\right)^{2}\left(m^2 + n^2 - in - \frac{\beta}{\gamma} + \frac{R^2}{H^2}\right)}}{ m^2 + n^2 - in - \frac{\beta}{\gamma} + \frac{R^2}{H^2}}\Omega.
\end{eqnarray}
In Fig.\ 5 the growth of the positive mode is plotted together with the integrated amplification
for a disk with $H/R = 0.1$, $m=1$, and $\beta = 1$ (solid line). 
The predicted amplification of $\approx$ 100 fits to the numerical result (section 5)
for the same parameters (see Fig.\ 8).
This confirms that our simplified dispersion relation yields useful results.
In addition we plotted the 
baroclinic solution for $\beta = 0$ (dotted line) to show that the important effect in
the baroclinic solution is the bifurcation of the solutions into a positive 
and negative solution branch. 

   The occurrence of this bifurcation depends on $\beta$. To study this bifurcation it
is sufficient to focus on the behavior of the solution at $n(t)=0$.
\begin{eqnarray}
\Gamma_l(0) &=& m \frac{i \left(\frac{5}{4}\frac{\beta}{\gamma} - 1 \right)}{ m^2 - \frac{\beta}{\gamma} + \frac{R^2}{H^2}}\Omega\\
&\pm& m \frac{\sqrt{ - \left(\frac{5}{4}\frac{\beta}{\gamma} - 1 \right)^2 + \frac{\beta^2}{\gamma^2}\left(\frac{H}{R}\right)^{2}\left(m^2  - \frac{\beta}{\gamma} + \frac{R^2}{H^2}\right)}}{ m^2 - \frac{\beta}{\gamma} + \frac{R^2}{H^2}}\Omega.
\end{eqnarray}
The first part is always imaginary 
determining the frequencies of the waves and
the second term leads only to a bifurcation of the modes in growth-rates, if the root has a real part.
Thus the condition for bifurcation and by that for transient growth is
\begin{eqnarray}
 - \left(\frac{5}{4}\frac{\beta}{\gamma} - 1 \right)^2 + \frac{\beta^2}{\gamma^2}\left(\frac{H}{R}\right)^{2}\left(m^2  - \frac{\beta}{\gamma} + \frac{R^2}{H^2}\right) > 0.
\end{eqnarray}
In general the critical $\beta$ for bifurcation is given by this third order polynomial. 
The dependence of $\Gamma$ on $\beta$ is 
plotted in Fig.\ \ref{fig6.ref} for the $m=60$ modes.
It appears that $m\sim R/H$ is required for maximum growth.

Basically there are bifurcating solutions for $\beta < -0.4$ which means a positive entropy gradient
and for $\beta > 0.25$.
The unstable regime for $\beta$ for modes with $m < R/H$ is given by
\begin{eqnarray}
 \frac{4}{9} \gamma < \beta < 4 \gamma
\end{eqnarray}
in the case of a radially constant mean density.
For larger $m$ (e.g.\ $m > \frac{H}{r}$) 
\begin{eqnarray}
  \beta > \frac{\gamma}{\left(\frac{H}{R}\right) m}
\end{eqnarray}
is valid. As a maximum reasonable azimuthal wave number one can assume $m = 2 \pi R / H$ because 
for larger wave numbers the vertical structure can not be neglected anymore.
Thus a critical minimum $\beta$ for the occurrence of a global baroclinic instability
should be $\beta_{min} = 0.22$. 

%
%
\section{Numerical Integration}
Non-axisymmetric solutions of Eqs.\ (41)-(44) with a 
time-dependence of the form $exp(i\omega t)$
do exist only in the absence of shear ($B=-\Omega$). 
To justify the usage of this simplified time dependence 
in a shearing system, as in section 3 and 4, one has to perform a numerical
test integration of Eqs.\ (41)-(44) without substituting the
time dependence with $exp(i\omega t)$, i.e.\
\begin{eqnarray}
\partial_t \Sigma_a  &=& - i n(t) v_{xa} - i m v_{ya},\\
\partial_t  v_{xa}   &=& - \beta \bar P \Sigma  + 2 v_{y a} - i n(t) P_a,\\
\partial_t  v_{ya} &=& - \frac{1}{2} v_{xa} - i m P_a  ,\\
 \partial_t  P_a   &=& - \left(i n(t) \gamma - \beta\right)  \bar P v_{xa} - i m \gamma \bar P  v_{y a},
\end{eqnarray}
with 
\begin{eqnarray}
n(t) = \frac{3}{2} m t
\end{eqnarray}
The integration is carried out using an implicit Runge-Kutta method (RADAU5) 
of order 5 with step size control (see Hairer \& Wanner 1996).
For the parameters, we assume $\Sigma_0 = 1.0, \Omega_0 = 1.0, R_0 = 1.0, m=60, H/R = 0.1,$ and $-40/\Omega_0 < t < 40/\Omega_0$
to compare the results to that presented in Fig.\ 5.
In Fig.\ 8, we plot the time evolution of the most unstable eigenfunction.
The highest frequencies with $\omega >> \Omega$ were properly resolved in the 
numerical integration, but they were removed from the all the following 
plots via a simple time filtering procedure. This does not change lower
frequencies.

The initial solution of the most unstable eigenfunction was found numerically for:
\begin{eqnarray}
n =  -40\frac{3}{2} m/\Omega_0 = -3600
\end{eqnarray}
by integrating the initial value problem $\Sigma_a(0)=\left(\sqrt{\frac{1}{2}}-i\sqrt{\frac{1}{2}}\right) \times 10^{-10}$ and 
$P_a(0)=v_xa(0)=v_ya(0)=0$ in time for the fixed $n = -3600$ and renormalising all amplitudes every couple of
time steps to keep the amplitude of the density fluctuation at $|\Sigma_a| = 10^{-10}$
until the solution converges.
As one can already see from Fig.\ 1, only one solution shows growth for large
negative wavenumbers. This is also the eigenfunction that shows the baroclinic
unstable mode eventually.
\subsection{Case A $\beta = 1$}
In Fig.\ 8 we plotted the results from the numerical integration of the initial perturbation
with $\beta = 1$. All values are scaled to the value of the initial 
surface density perturbation amplitude.
The surface density perturbation has qualitatively the same behavior as
in the analytic estimate (compare with fig.\ 5). The quantitative behavior is different
in a sense that the swing amplification factor is about two times larger then
estimated. On the other hand, the initial amplification of the pressure and velocity fluctuations 
vanishes almost completely for large times. 
We also plot the amplitude of the potential vorticity perturbation $q$:
\begin{eqnarray}
q(t) = \frac{\omega(t) + \frac{1}{2}{\Omega}}{\Sigma_0 + \Sigma_a(t)} - \frac{1}{2}{\Omega}
\end{eqnarray}
Here $\omega(t)$ is the vorticity:
\begin{eqnarray}
\omega(t) = \nabla \times v_a(t)  = i \left[k_x(t) v_{ya}(t) - k_y v_{xa}(t)\right],
\end{eqnarray} and must not to be confused with the frequencies of the eigenfunction.
Potential vorticity (also called vortensity) 
is a conserved quantity in barotropic flows.
But in baroclinic flows it can be generated,
which can be seen in Fig.\ 8. The amplification in potential vorticity is
not very dramatic compared to the initial state. Later in Case C (Fig.\ 10), we perform a calculation with vortensity generation starting from an initial state of
zero local vorticity.

   Right after $t=0$, high frequency waves show up in
the amplitudes. These are the sound waves which are unfortunately
also part of the solution. But their frequency diverges and the
plotting routine filters them out successfully.
\subsection{Case B $\beta = 0$}
If one integrates the equations using the same initial values 
from last section 
but now with $\beta = 0$, the situation looks very similar
for times $t<0$ (see Fig. 9). This fits perfectly to the expectations from
Fig.\ 5. In contrast to Fig.\ 8, now all amplitudes decay for times $t>0$.
This also agrees with the analytic estimates.

   In addition, one can now observe that vortensity is conserved, 
as it should be the case for barotropic flows.

\subsection{Case C $\omega(t=0) = 0$}
For the numerical integrations in Fig.\ 10, we did not use
the numerically derived most unstable eigenfunction but an
eigenfunction that contains initially zero vorticity ($\Sigma_a(0)=\left(\sqrt{\frac{1}{2}}-i\sqrt{\frac{1}{2}}\right) \times 10^{-10};$ and $ P_a(0)=v_xa(0)=v_ya(0)=0$). Thus, we can show how effectively potential vorticity is generated in the
baroclinic case $\beta = 1$ in comparison to the barotropic situation, where $\omega = 0$ at all times. This time, surface density is not amplified by as
much as in the case of the pure eigen function (case A).

We conclude that the usage of the simplified time dependence $(\sim exp(i\omega t))$
for the derivations of the eigenvalues of the problem and the critical parameters for stability and instability is justified (see also Ryu \& Goodman 1992). 
\subsection{Case D $\beta = -0.5$}
For completeness we also added a case with a negative $\beta$, which corresponds
to a positive pressure (entropy) gradient.
As expected, the result (see Fig.\ 11) looks very similar to 
Case A, only the potential vorticity generation is smaller this time.
Thus, we stress again: disks with any pressure gradient (negative and
positive) can undergo this baroclinic instability.
%
%
\section{Explanation of the fundamental instability cycle}
The linearized equations in the shearing sheet formalism allow a simple 
study of the basic instability cycle. 
Therefore, one neglects the radial derivative  
$\partial_x << \partial_y$ and focuses on times close to $t=0$,
where the effect
(and the bifurcation) is the clearest:
\begin{eqnarray}
\partial_t \Sigma^\prime  &=& - \Sigma_0 \partial_y v_y^\prime, \label{simple_cont}\\
\partial_t v_x^\prime &=& - \frac{\Sigma^\prime}{\Sigma_0^2} \frac{\beta}{R} \bar{P} + 2 \Omega v_y^\prime,\label{simple_vx}\\
\partial_t v_y^\prime  &=& - \frac{1}{\Sigma_0}\partial_y P^\prime + 2 B v_x^\prime,\label{simple_vy}\\
\partial_t P^\prime  &=& v_x^\prime \frac{\beta}{R} \bar{P}-\gamma \bar{P} \partial_y v_y^\prime.\label{simple_p}
\end{eqnarray}
A simple sketch will elucidate what happens in 
the disk (see Fig.\ \ref{fig7.ref}).
If there is an azimuthally confined over-density in the disk$\Sigma^\prime > 0$,
but no other perturbation (especially no perturbation of the pressure) ,
then equation \ref{simple_vx} predicts a radial inward drift of matter,
\begin{equation}
\partial_t {v_x}^\prime = - \frac{\beta}{r} P_0  \frac{\Sigma^\prime}{\Sigma_0^2}\,\, => \,\,{v_x}^\prime < 0,
\end{equation}
because the disk is no longer in radial equilibrium. This is 
basically radial buoyancy in a rotating system.
A region with lower density but constant pressure must have a higher
temperature.
The radial inward drift $v_x < 0$ then decreases the local pressure as follows from Eq.\ \ref{simple_p}:
\begin{equation}
\partial_t P^\prime =  \frac{\beta}{r} P_0 v_x^\prime\,\, =>\,\, {P}^\prime < 0.
\end{equation}
But, this will result in an azimuthal velocity towards the pressure minimum (see Eq.\ \ref{simple_vy}):
\begin{equation}
\partial_t v_y^\prime = - \frac{1}{\Sigma_0}\partial_y P^\prime\,\, => \,\,\partial_y v_y^\prime< 0
\end{equation}
which in the end will lead to a further increase of azimuthally concentrated 
density (see Eq.\ \ref{simple_cont}):
\begin{equation}
\partial_t {\Sigma}^\prime = - \Sigma_0 \partial_x {v_x}^\prime\,\, =>\,\, \Sigma^\prime > 0.
\end{equation}
Now the cycle is closed, and a further amplification can follow.

%
%

\section{Discussion and Conclusions}

\subsection{Do Accretion disks have a negative Entropy gradient $\beta > 0.22$ ?}
Isothermal disks have no entropy gradient and will be stable. But, not every disk
can be assumed to be isothermal.
Three scenarios in which the instability criterion can be fulfilled are given here:
\begin{itemize}
\item{A Disk close to the central object or to the boundary layer:}\\
In such a situation, the temperature gradient is determined by the radial radiation transport in the disk.
The temperature profile can be very steep in an optical thick disk, irradiated only at the inner edge.
\item{A protostellar disk:}\\
A disk that has just formed after the collapse of
a molecular cloud core is also likely to have a
temperature gradient slightly steeper than 0.5 (see Fig.\ 4. in Yorke \& Bodenheimer 1999).
\item{A disk with self sustained accretion:}\\
Under the assumption of a constant accretion rate, the locally produced energy 
gives a temperature profile of $T \sim r^{-0.75}$, which will be unstable to 
the Global Baroclinic Instability.
Two- and three-dimensional radiation hydro simulations of accretion disks showed a temperature
slope of $T \sim r^{-1}$ while the surface density was constant over wide ranges of radii at $3.5-6.5$ AU
(Klahr \& Bodenheimer 2003).
\end{itemize}
Detailed studies on the actual entropy gradient in various disks will have to follow this study.

   So far only the simple case of constant surface 
density was studied in the linear analysis, which is indeed the proper solution for
an $\alpha$-model with dust opacities for a constant
accretion rate at about $5 \pm 3$AU (see Bell, Cassen, Klahr, \& Henning 1997).

Nevertheless, an analysis for general $\beta_\Sigma$ will have to follow.
What one can expect from this more general analysis is a stability criterion that not only depends on the entropy gradient
but the entropy gradient in relation to the local pressure gradient in the fashion of the Schwarzschild-Ledoux criterion:
\begin{equation}
\frac{dP}{dr}\frac{dS}{dr} > 0
\end{equation}
Only when entropy gradient and pressure gradient point in the same direction a convective instability occurs.
As long as the surface density is radially constant, this criterion is always fullfilled.

\subsection{Closing remarks}
Rotating centrifugally supported systems (e.g., accretion disks, galaxies, fast rotating stars) 
can be hydrodynamically unstable
to non-axisymmetric perturbations under the influence of a radial entropy gradient. 
Even if they are stable with respect to the Rayleigh and S{\o}lberg-Hoiland
criterion, the radial entropy gradient can produce a Baroclinic instability forming slowly oscillating
waves. A similar effect was known in the case of planetary surfaces (Classical Baroclinic Instability) 
and has now been demonstrated for accretion disks.

   While the instability produces linear growth for
rigidly rotating bodies, which is known since the 1940s (e.g.\ Cowling 1951), it is weakened by the
radial shear inherent in accretion disks. 
Thus, in sheared geometries, the
instability is transient, leading only to a finite amplification in
the linear regime. Non-linear effects can then possibly pick up
and lead to further amplification.

It could be possible that the nonlinear vortices arise through the
following three steps.
First, swing amplification of an initially
nonvortical leading perturbation produces a vortical trailing one.  We
know that in a barotropic disk, a perturbation in vorticity per unit
surface density is permanent (absent dissipation) once it is created.
In these linearized calculations, one expects that the trailing
vortical perturbations will be approximately constant in amplitude at
late times because the baroclinic source terms for vortensity
become small when the wave is tightly wrapped 
(large $k_x(t)/k_y$).  There will then
be an almost axisymmetric pattern of radial maxima and minima in
vortensity, which gradually become more pronounced as $k_x(t)$ increases.
Thus, the second step may be a Kelvin-Helmholtz instability of this pattern,
resulting in small-scale vortices (``Kelvin's cat's-eyes'').  The third
and final step would be the merging of small anticyclonic vortices into
larger ones. 

This would explain the results
from non-linear simulations (Klahr \& Bodenheimer 2003) and also
give a physical model for the vortex formation in protoplanetary disks,
which is supposed to be the starting point of planet formation (Klahr 2003).
It would be useful to make predictions for the shape, size, and frequency
of these large-scale anti-cyclonic vortices before they become observable
with interferometric instruments like ALMA (Wolf \& Klahr 2002).

\acknowledgments
My special thanks to Steve Balbus, Peter Bodenheimer, Ethan
Vishniac, Willy Kley, James Cho, and most of all the referee for
very useful criticism and helpful suggestions.
This work was in part supported by the NASA grant NAG5-4610 and by a
special NASA astrophysics theory program which supports a joint
Center for Star Formation Studies at NASA-Ames Research Center, UC
Berkeley, and UC Santa Cruz.

\clearpage





\clearpage
\begin{figure}
\plotone{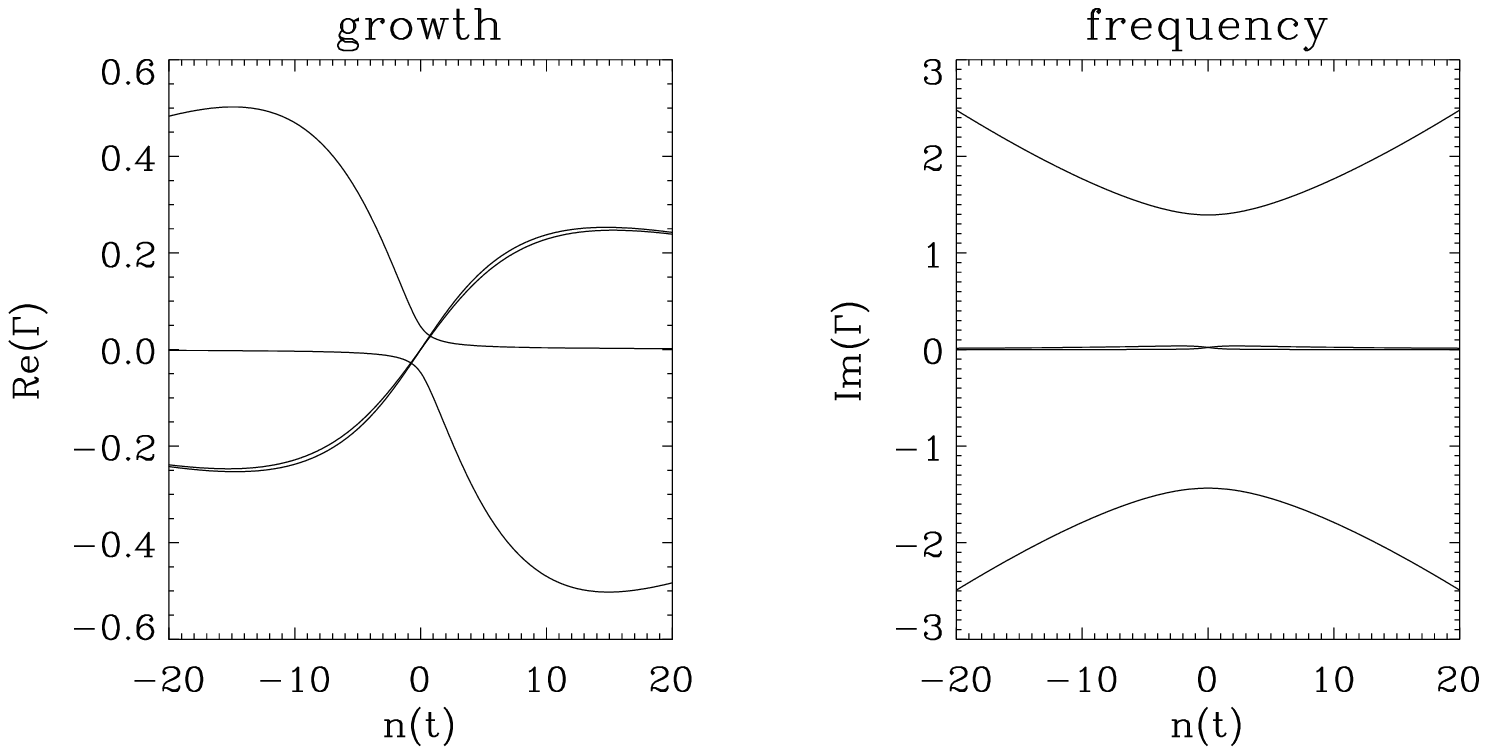}
\caption{Solutions of the local dispersion relation: Growth rates and frequencies of the $m = 10$ modes as a function of the time-dependent radial wave number $n(t) = n_0 + 3/2 \Omega m t$  in a narrow range
from $-20 < n < 20$.
For explanation see text.}
\label{fig1.ref}
\end{figure}
\clearpage
\begin{figure}
\plotone{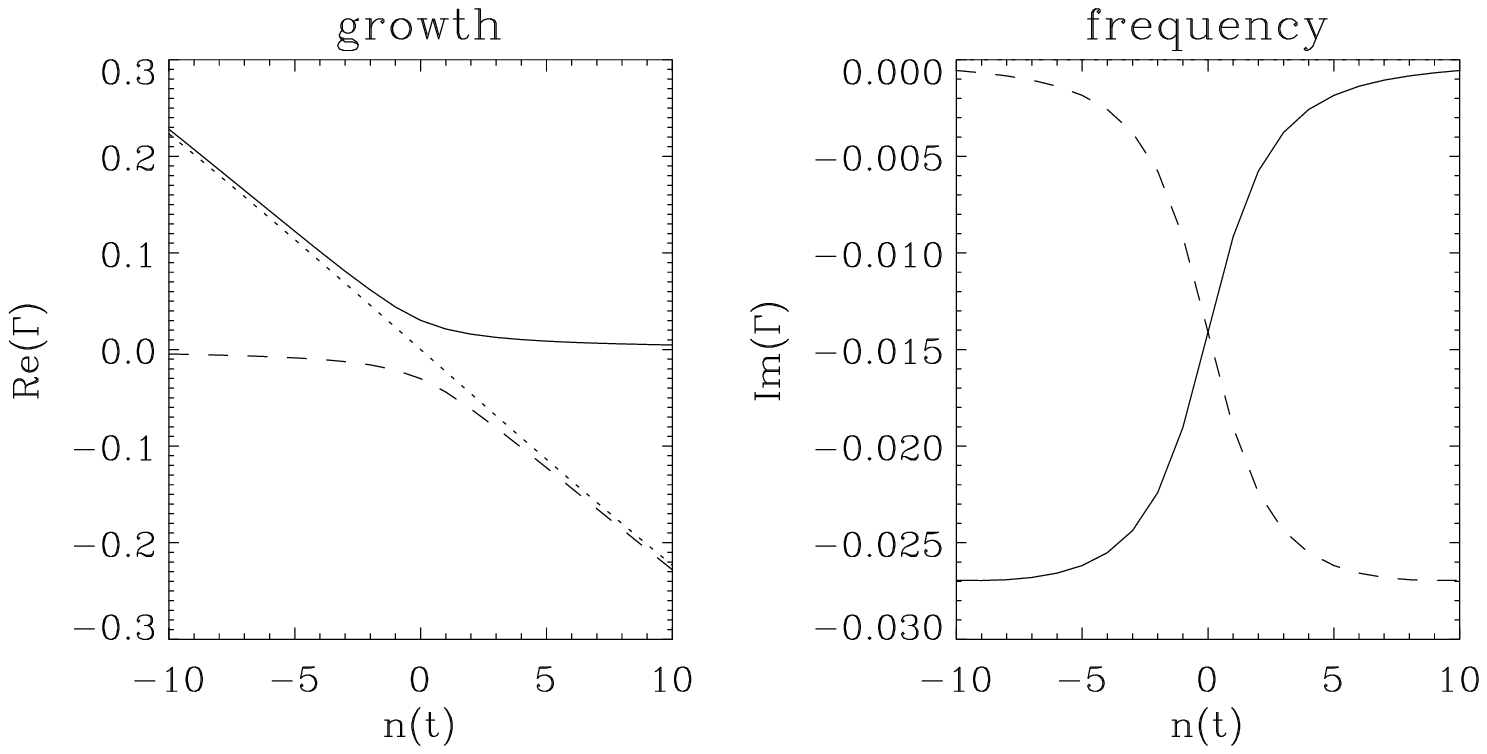}
\caption{Solutions of the local dispersion relation for $\beta = 1$: Growth rates and frequencies of the $m = 48$ modes as a function of the timedependent radial wave number $n(t) = n_0 + 3/2 \Omega m t$  in a narrow range
from $-10 < n < 10$.
The dotted line is the barotropic case $\Gamma_0$, while the baroclinic case
is given by the solid $\Gamma^+$ and dashed $\Gamma^-$ curve.
For explanation see text.}
\label{fig2.ref}
\end{figure}

\begin{figure}
\plotone{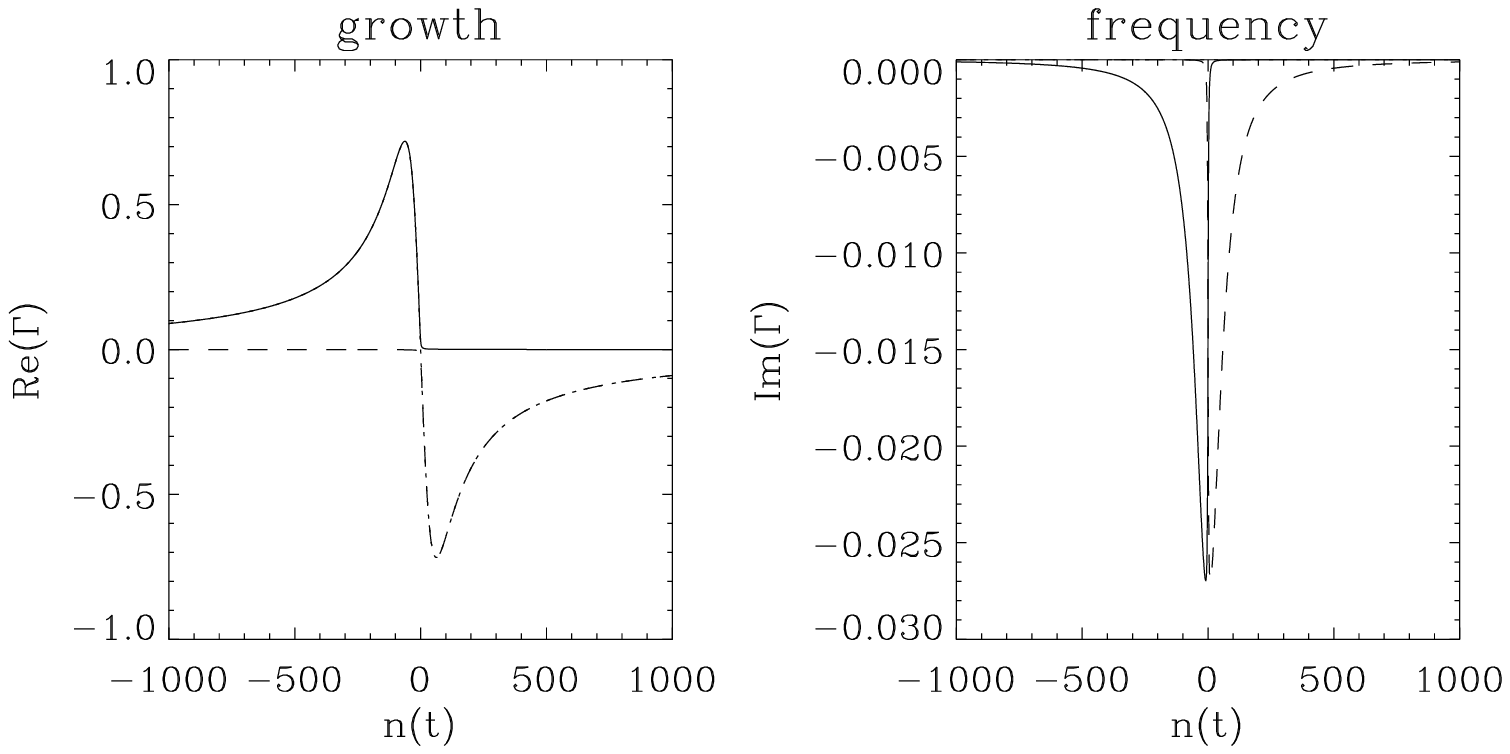}
\caption{Solutions of the local dispersion relation for $\beta = 1$: Growth rates and frequencies of the $m = 48$ modes as a function of the timedependent radial wave number $n(t) = n_0 + 3/2 \Omega m t$ in a wide range
from $-1000 < n < 1000$.
The dotted line is the barotropic case $\Gamma_0$, while the baroclinic case
is given by the solid $\Gamma^+$ and dashed $\Gamma^-$ curve. Same data as Fig.\ 2.
For explanation see text.}
\label{fig3.ref}
\end{figure}
\clearpage

\begin{figure}
\plotone{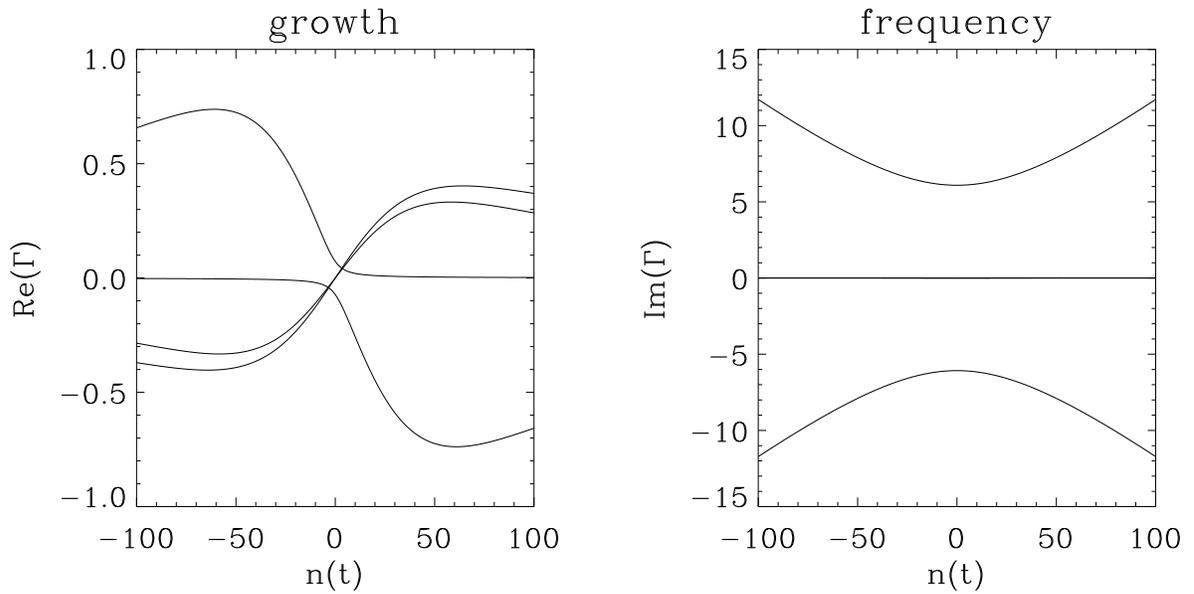}
\caption{Solutions of the global dispersion relation for $\beta = 1$: Growth rates and frequencies of the $m = 60$ modes as a function of the time-dependent radial wave number $n(t) = 3/2 \Omega m t$  in the range
from $-40 < n < 40$.}
\label{fig4.ref}
\end{figure}

\begin{figure}
\plotone{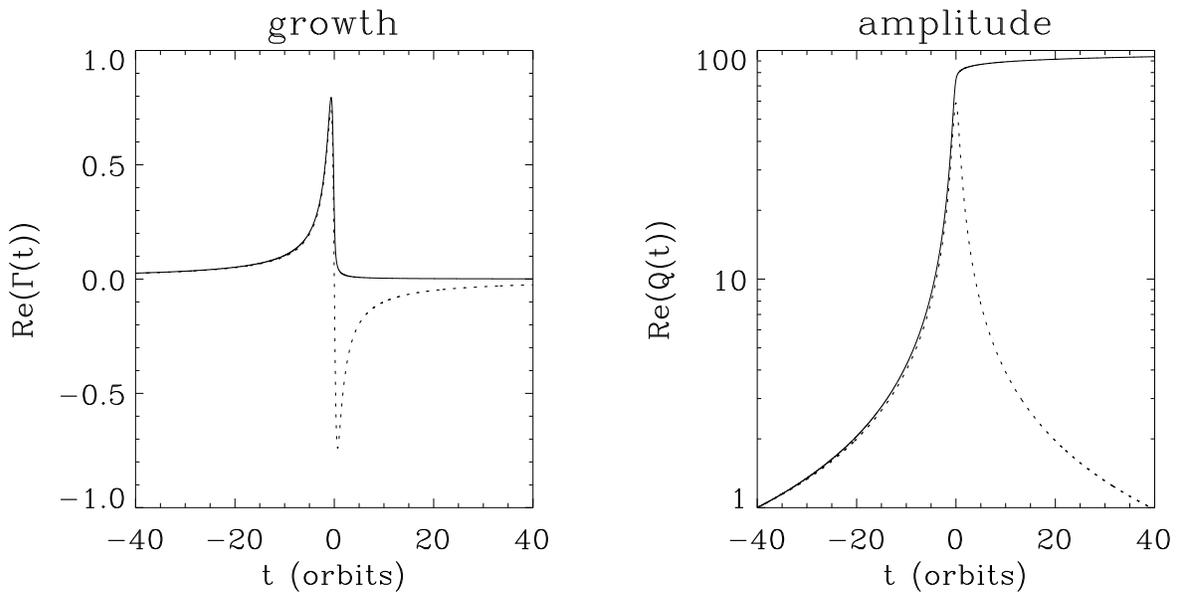}
\caption{The transient unstable mode of the global dispersion relation: Growth rates $\Gamma^+$ and integrated amplification $Q(t)$ of the $m = 60$ mode as a function of time in units of orbital periods. Compare the baroclinic case ($\beta = 1$: solid line) with the barotropic case $\Gamma_0$ ($\beta = 0$: dotted line). (Unlike in the previous figures
time $t$ is used instead of $n(t) = 3/2 \Omega m t$ for this plot.)}
\label{fig5.ref}
\end{figure}

\begin{figure}
\plotone{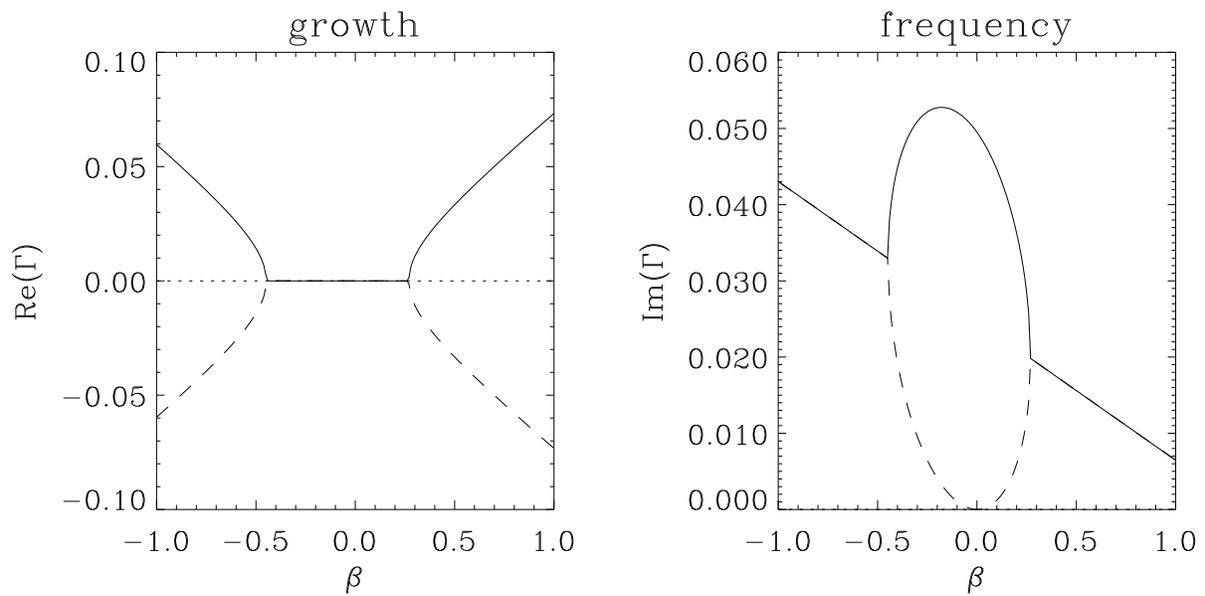}
\caption{Solutions of the global dispersion relation: Growth rates and frequencies of the $m = 60$ modes at time $t = 0$ as a function of the 
radial pressure gradient $\beta$. Bifurcation in growth-rates only occurs for
certain values of $\beta$.
The dotted line is the stable $m=0$ case, while the baroclinic case
is given by the solid $\Gamma^+$ and dashed $\Gamma^-$ curve.
For explanation see text.}
\label{fig6.ref}
\end{figure}

\begin{figure}
\plotone{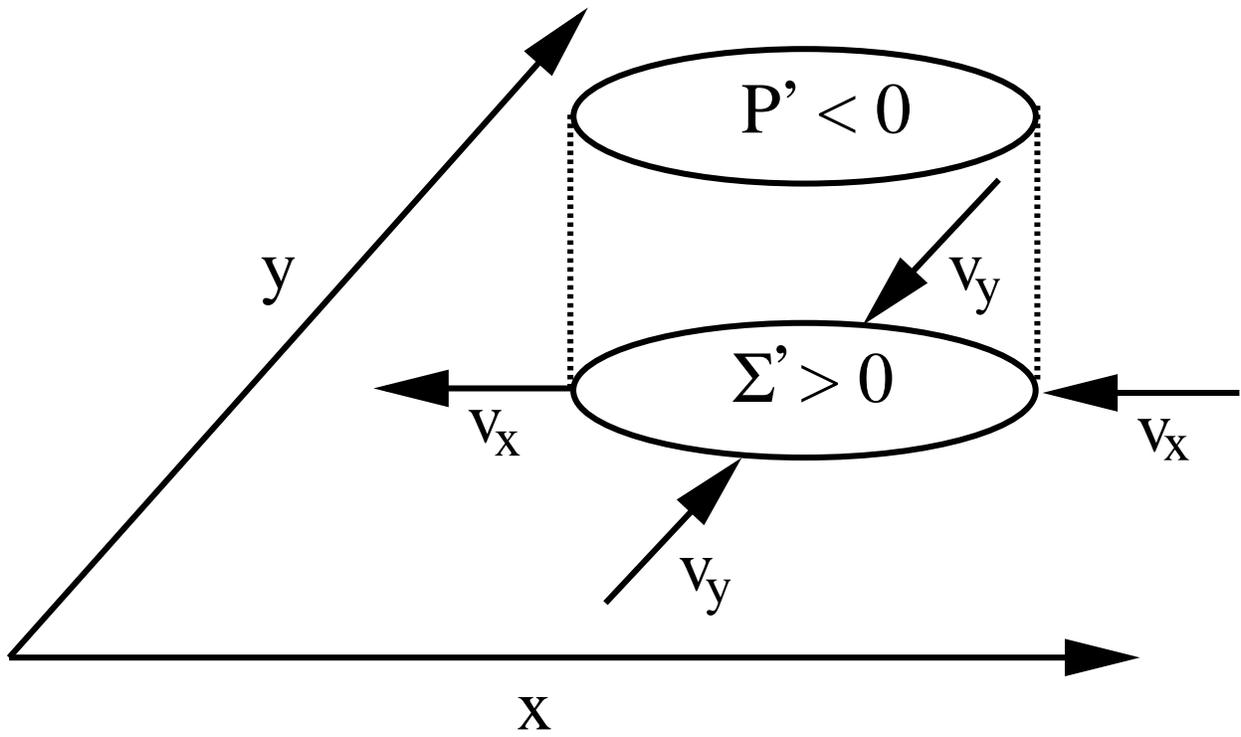}
\caption{Schematic process of the Instability cycle. For explanation see text.} 
\label{fig7.ref}
\end{figure}

\begin{figure}
\plotone{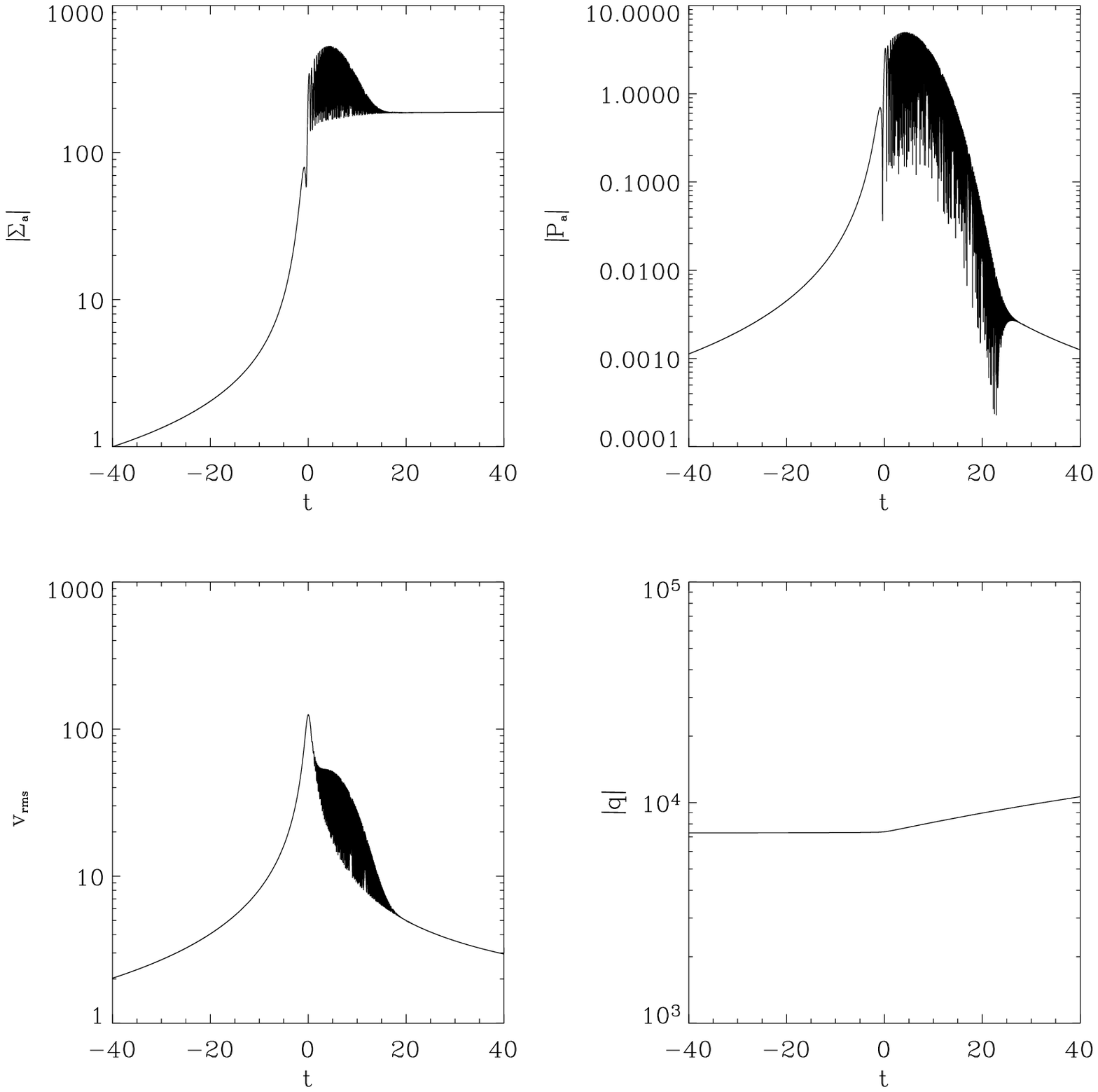}
\caption{Evolution of the perturbations in a Keplerian disk for Case A ($\beta=1$):
The amplitudes of surface density $\Sigma_a$, pressure $P_a$, root mean square
velocity $v_{rms}$ and potential vorticity $q$ are plotted. For explanation see text.}
\label{fig8.ref}
\end{figure}

\begin{figure}
\plotone{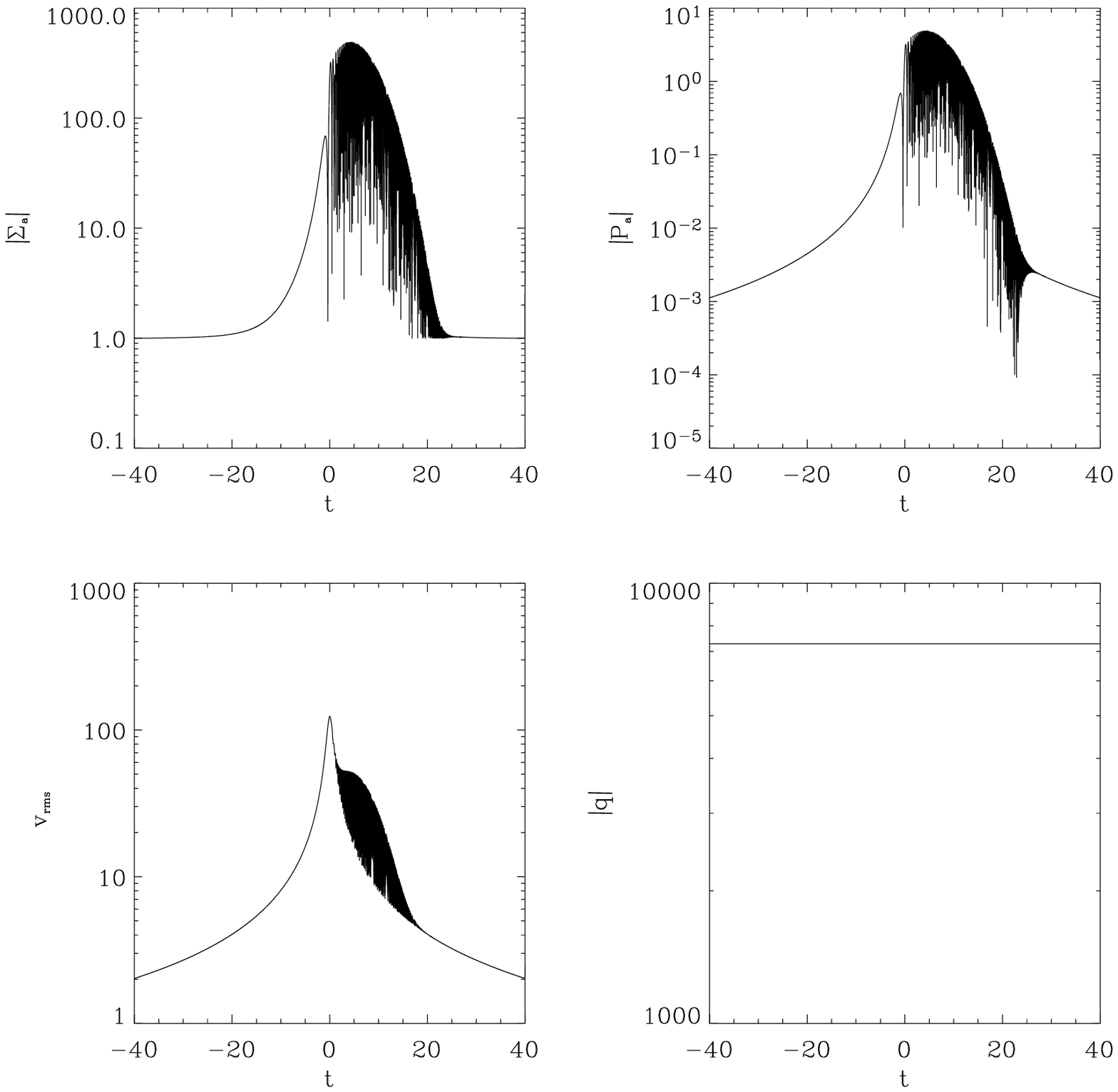}
\caption{Evolution of the perturbations in a Keplerian disk for Case B ($\beta=0$):
The amplitudes of surface density $\Sigma_a$, pressure $P_a$, root mean square
velocity $v_{rms}$ and potential vorticity $q$ are plotted. For explanation see text.}
\label{fig9.ref}
\end{figure}

\begin{figure}
\plotone{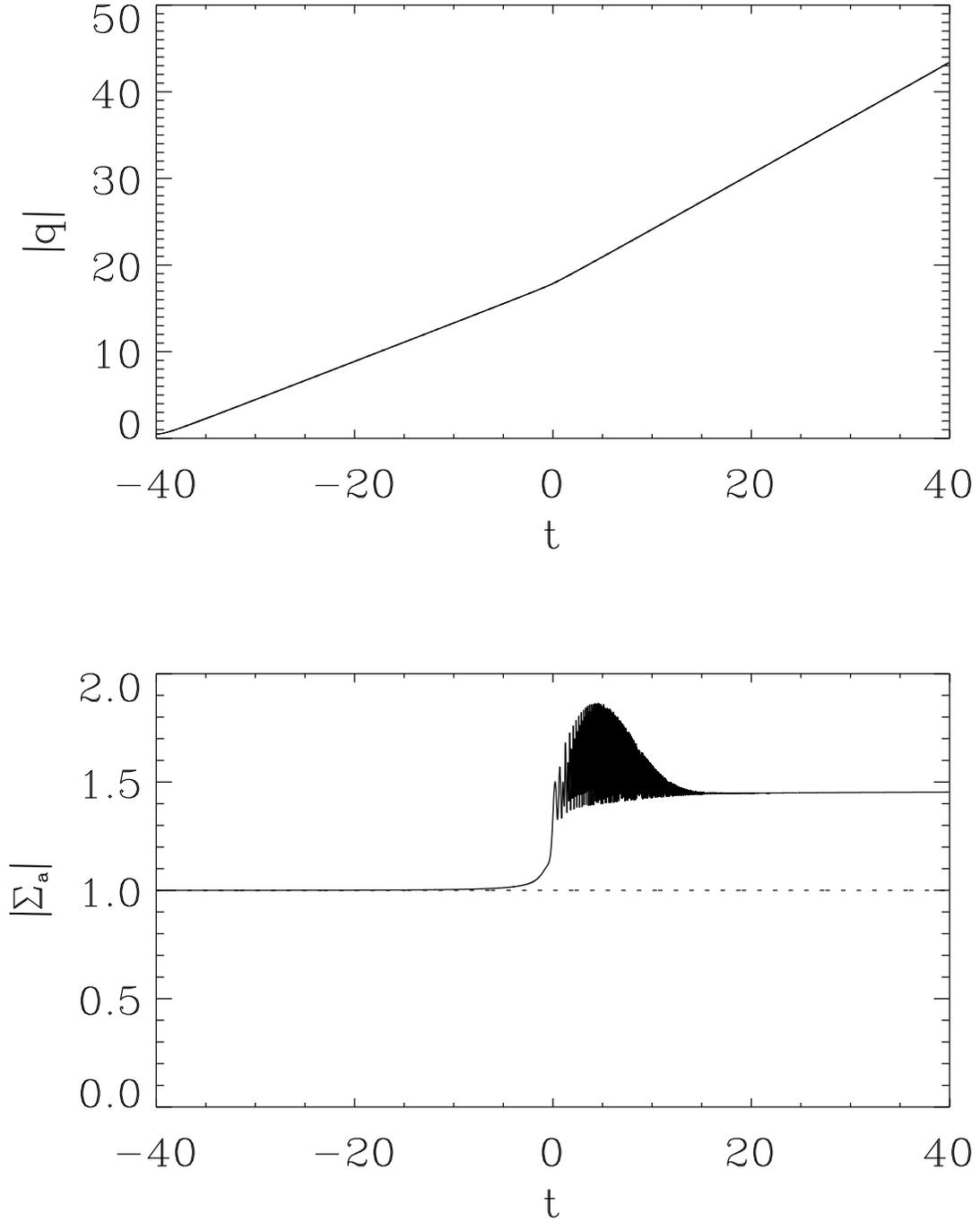}
\caption{Evolution of the perturbations in a Keplerian disk for Case C ($\beta=1$
and zero initial vorticity):
The amplitudes of surface density $\Sigma_a$ and potential vorticity $q$ are plotted. 
For explanation see text.}
\label{fig10.ref}
\end{figure}

\begin{figure}
\plotone{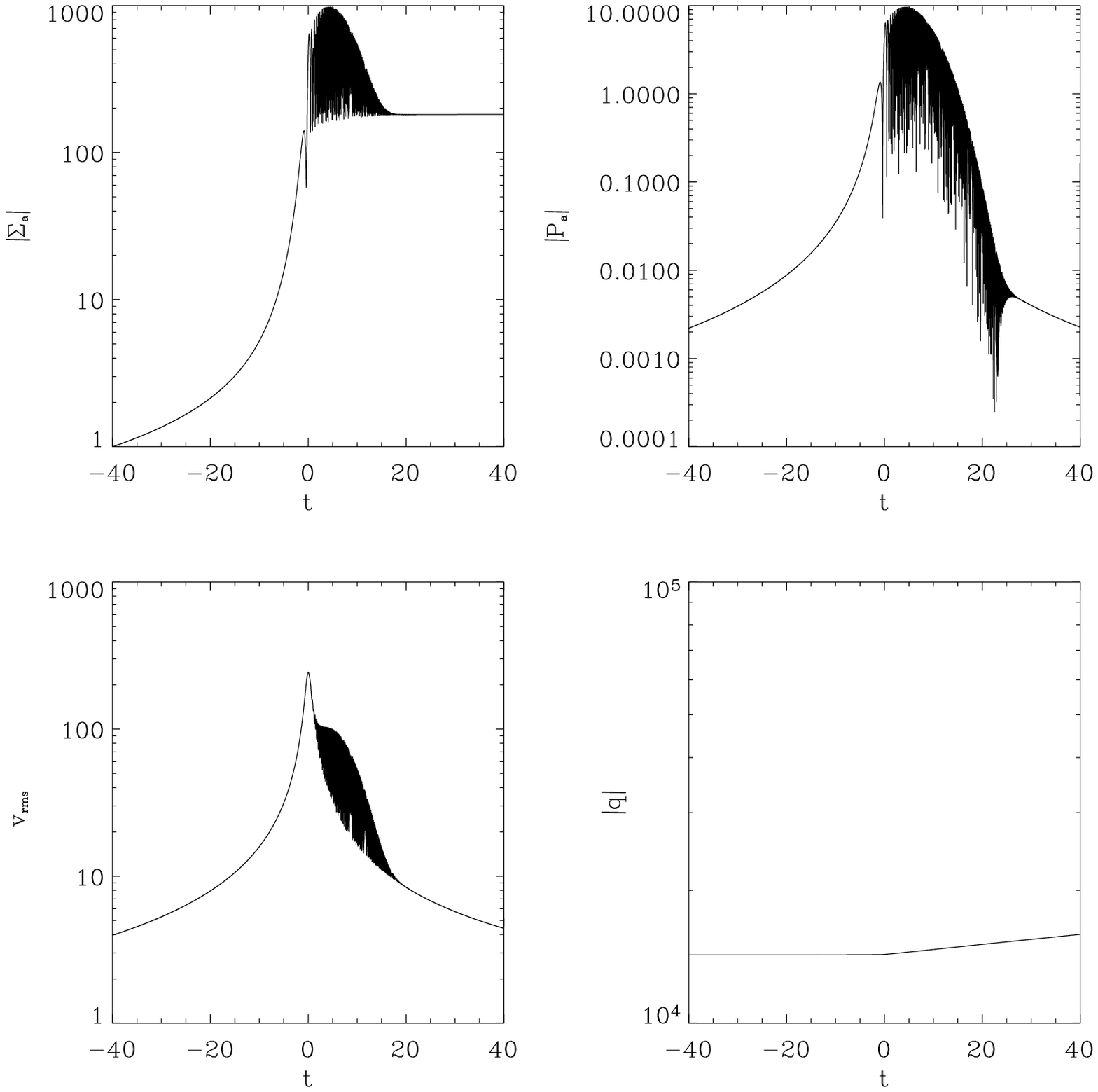}
\caption{Evolution of the perturbations in a Keplerian disk for Case D ($\beta=-0.5$):
The amplitudes of surface density $\Sigma_a$, pressure $P_a$, root mean square
velocity $v_{rms}$ and potetntial vorticity $q$ are plotted. For explanation see text.}
\label{fig11.ref}
\end{figure}


\begin{thebibliography}{}
\bibitem[]{} Balbus, S.A., \& Hawley, J.F.\  1991, \apj, 376, 214
\bibitem[]{} Balbus, S.A., \& Hawley, J.F.\  1992, \apj, 400, 610
\bibitem[]{} Balbus, S.A., Hawley, J.F., \&  Stone, J.M.\ 1996, \apj, 467, 76
\bibitem[]{bell97} Bell, K.R., Cassen, P.M., Klahr, H., \& Henning, Th.\ 1997, \apj, 486, 372
\bibitem[]{} Binney, J.\& Tremaine, S.\ 1987, Galactic dynamics (Princeton, NJ, Princeton University Press)
\bibitem[Cowling(1951)]{1951ApJ...114..272C} Cowling, T.~G.\ 1951, \apj, 
114, 272
\bibitem[]{} Gammie, C.F.\ 1996, \apj, 457, 355
\bibitem[]{} Goldreich, P., Goodman, J., \& Narayan, R.\ 1986, \mnras, 221, 339
\bibitem[]{} Goldreich, P., \& Lynden-Bell, D.\ 1965, \mnras, 130, 125
\bibitem[]{} Hairer, E., \& Wanner, G.\ Springer Series in Computational Mathematics 14,
         Springer-Verlag 1991, 2nd.\ ed.\ 1996
\bibitem[]{} Klahr, H.,  \& Bodenheimer, P.\ 2000, 
Proceedings of: Disks, Planetesimals and Planets, ASP Conference Series, eds. F. Garzón, C. Eiroa, D. de Winter, \& T. J. Mahoney (Astronomical Society of the Pacific) Vol. 219, 63
\bibitem[]{} Klahr, H.,  \& Bodenheimer, P.\ 2003,  \apj, 582, 869
\bibitem[]{} Klahr, H.\ 2003, Proceedings of: Scientific Frontiers in Research on Extrasolar Planets, ASP Conference Series, eds. D. Deming, \& S.\ Seager (Astronomical Society of the Pacific) Vol.\ 294, 277
\bibitem[]{} Pedlosky, J.P.\ 1987, Geophysical Fluid Dynamics, 2nd ed. (New York: 
Springer-Verlag)
\bibitem[]{}R\"udiger, G., Arlt, R., \& Shalybkov, D.\ 2002, \aap, 391, 781
\bibitem[]{}Ryu, D., \& Goodman, J.\ 1992, \apj, 388, 438
\bibitem[]{}Toomre, A.\ 1964, \apj, 139, 1217
\bibitem[]{wo99} Wolf, S., \& Klahr, H.\ 2002, \apj, 578, L79
\bibitem[]{}Yorke, H.\ \& Bodenheimer, P.\ 1999 \apj, 525, 330
\end{thebibliography}
\end{document}